# Partial wavelet coherence analysis for understanding the standalone relationship between Indian Precipitation and Teleconnection patterns

Maheswaran Rathinasamy[1,] Ankit Agarwal[2, 3,*,] Vilakshna Parmar[4], Rakesh Khosa[4] and Arvind Bairwa[4]

[1]MVGR College of Engineering, Vizianagaram, India
[2] Institute of Earth and Environmental Science, University of Potsdam, Potsdam, Germany
[3]Potsdam Institute for Climate Impact Research, Telegrafenberg, Potsdam, Germany
[4]Indian Institute of Technology Delhi, New Delhi

*Corresponding author at: Institute of earth and environmental sciences, University of Potsdam, Germany, Tel: +49 331 977 5433, fax: +49 331 977 2092
Email address: aagarwal@uni-potsdam.de (A. Agarwal)

## Abstract

Hydro-meteorological variables, like precipitation, streamflow are significantly influenced by various climatic factors and large-scale atmospheric circulation patterns. Efficient water resources management requires an understanding of the effects of climate indices on the accurate predictability of precipitation. This study aims at understanding the standalone teleconnection between precipitation across India and the four climate indices, namely, Niño 3.4, PDO, SOI, and IOD using partial wavelet analysis. The analysis considers the cross correlation between the climate indices while estimating the relationship with precipitation. Previous studies have overlooked the interdependence between these climate indices while analysing their effect on precipitation. The results of the study reveal that precipitation is only affected by Niño 3.4 and IOD and a non-stationary relationship exists between precipitation and these two climate indices. Further, partial wavelet analysis revealed that SOI and PDO do not significantly affect precipitation, but seems the other way because of their interdependence on Niño 3.4. It was observed that partial wavelet analysis strongly revealed the standalone relationship of climatic factors with precipitation after eliminating other potential factors.

**Abbreviations**

ENSO: El Nino Southern Oscillation

PDO: Pacific Decadal Oscillation

SOI: Southern Oscillation Index

IOD: Indian Ocean Dipole

SST: Sea surface temperature

1.  **Introduction**

Efficient water resource management requires an understanding of the effects of natural climate variability on precipitation, particularly in the context of increasing climate uncertainties. Analysis of the precipitation records is very important in understanding its variability and the underlying driving forces of natural systems. The task of quantifying precipitation variability, particularly as it relates to climate indices through teleconnections, has been explored in numerous studies (Simpson and Colodner (1999); Bonsal et al. (2001); Enfield et al. (2001); Gray (2004); Coulibaly and Burn (2004); Soukup et al. (2009); Rajagopalan et al. (2000); Özger et al. (2009), Valdivia et al. (2012), Malik et al. (2012); Marwan and Kurths (2015)).

In recent years, wavelet analysis has been increasingly used for analyzing highly irregular, complex, and intermittent nonstationary time series often encountered in geophysics. Further, wavelet analysis has been recognized as a useful technique for estimating the teleconnections between different climate indices and hydro-meteorological variables. Owing to the advantages of wavelet transforms, techniques such as cross-wavelet analysis (CWT) and wavelet coherence (WC) have also emerged as powerful tools in testing possible linkages between two signals (Grinsted et al. (2004)). Labat (2005) has provided a comprehensive review of wavelet based techniques like wavelet coherence, cross wavelet transform and wavelet modulus maxima (WMM) for applications in geophysical time series.

Studies such as Gan et al. (2007) investigated the teleconnections between Canadian precipitation and climate anomalies using wavelet coherence analysis. Mokhov et al. (2011)

analysed the relationship between ENSO and Indian precipitation using cross-wavelet analysis and Granger causality estimation from empirical data for the period of 1871–2003. Holman et al. (2011) studied the relationship between the groundwater levels in the UK and North Atlantic Oscillation (NAO) using the cross-wavelet analysis. Zhang et al. (2007) investigated the possible influence of ENSO on annual maximum streamflow pattern of Yangtze River, China and concluded that the stream flow observations pertaining to the middle and lower Yangtze River were dominated by 2- to 8-year periods. Similarly, Schaefli et al. (2007) presented a comprehensive review of applications of all wavelet methods such as wavelet spectrum and cross-wavelet spectrum to daily discharge, temperature, and precipitation.

Aforementioned studies have utilized wavelet coherence analysis in understanding the effect of teleconnections on hydrological variables. However, It needs to be highlighted that these studies do not consider the cross correlation between the teleconnections themselves (Gan et al. (2007) and He et al. (2014)) in estimating their influence on precipitation For example, the observable of precipitation (P) might be influenced by two variables $X_1$ and $X_2$ such that the two variables may themselves be correlated. In this situation, analysis of the relationship between P and $X_1$ using the wavelet coherence analysis inherently considers the effect of $X_2$. This might result in an erroneous interpretation of the actual relationship between P and $X_1$. Therefore, it is important to study the standalone effect of every climate index on precipitation distinctly, for making reliable predictions. However, it is not possible in an analytical way to study the standalone effect of the climate indices, however in a recent study, Mihanovic (2009) has presented the concept of partial wavelet coherence (PWC), providing a statistical way to estimate the standalone dependence of the two variables after eliminating the effect of other potentially influencing variables.

The objective of this paper is to use the concept of partial wavelet coherence analysis in understanding the standalone dependence between a given climate index and precipitation as it may shed more light on understanding the underlying natural processes. We hypothesise that disentangling the effects of the interrelationship of variables will improve the understanding of the influence of climate indices on Indian precipitation.

The present study utilized linear wavelet analysis to comprehend the dependence between climatic factors and precipitation. Nonetheless, as the future extension, one can utilize nonlinear dependence measures based on mutual information theory to investigate the presence of nonlinear relationships.

The standalone relationship between temporal components of rainfall anomaly and climate signals at different time scales have not been reported so far in the literature as per author's best knowledge.

## 2. Study Area and data

India has been divided into 36 meteorological sub-divisions (34 on the mainland and 2 on islands) (Figure 1a) out of which 30 meteorological subdivisions were utilized in this study (Kumar et al. (2010)) based on data availability. Figure 1b demonstrates all the 30 sub-divisional precipitation stations chosen in this study on the terrain India. Sub-divisional monthly rainfall data prepared by the Indian Institute of Tropical Meteorology (IITM) is used for the period of 1901- 2002 (source: http://www.india-wris.nrsc.gov.in/wris.html). Notably, the quality checks on the data were made to ensure that an error free data is made available for analysis and design.

Further, on the basis of regional homogeneity, India has been classified into 6 regions (Southern Peninsular, North East, North West, Central North East, North and Central India) by IITM is shown in Figure 1b. In this study, five to six stations from each of these homogeneous regions are considered as representative station and a detailed analysis was done.

**Climate index data:**

In general, precipitation is tele-connected to several types of climate indices and the knowledge of teleconnection pattern and strength of the interrelationship gives some amount of predictability in remote locations (sometimes as long as a few seasons). For instance, predicting El Niño enables prediction of North American rainfall, snowfall, droughts or temperature patterns with a few weeks to months' lead time (Gan et al. (2007)). In the context of the Indian Precipitation, many different climate indices (SOI, Nino 3.4, NAO, AMO, IOD, and PDO) have been shown to have teleconnections with precipitation patterns (reference). The following section provides a brief summary of the climate indices considered in this study.

a) IOD Index is represented by the by anomalous SST gradient between the western equatorial Indian Ocean (50°E–70°E and 10°S– 10°N) and the southeastern equatorial Indian Ocean (90°E–110°E and 10°S–0°N). It has been shown in several studies that IOD plays a key role in the climate of the Indian subcontinent. The data for IOD was obtained from the Bureau of Meteorology, Australia.

b) The Niño 3.4 index is one of the different indexes to measure the ENSO effect. Nino 3.4 is estimated as the average sea surface temperature anomaly in the region bounded by 5°N to 5°S, from 170°W to 120°W. Nagesh and Maity(2006) and several other have shown Nino 3.4 to have some influence on Indian monsoon.

c) Pacific Decadal Oscillation (PDO) is an ocean atmospheric climate index which is recurring over the mid latitude Pacific. Krishnamurthy and Krishnamurthy (2013) showed that the Indian monsoon rainfall decadal oscillations were shown to be associated with the decadal variability of the PDO.

d) North Atlantic Oscillation (NAO): It is calculated as the normalized pressure difference between the Azores and a station in Iceland.

e) American Multi decadal Oscillation (AMO): It is measured as the average anomalies of sea surface temperatures (SST) in the North Atlantic basin, typically over 0-80N.

f) Southern Oscillation Index (SOI) measures the difference in surface air pressure between Tahiti and Darwin. It is one of the key climate indices measuring the El Niño and La Niña events. The data for Niño 3.4, NAO, AMO, PDO, and SOI was downloaded from http://www.esrl.noaa.gov/psd/data/climateindices website

Although both, SOI and Nino.3.4, measure the same phenomenon of El-Nina and La Nina events yet there is currently no consensus in the scientific community as to which of these indices best capture ENSO phases. Therefore, in this study, we have used both the indices for a detailed analysis.

**Homogeneity test**

Before embarking on the application of wavelet analysis, the precipitation time series were tested for homogeneity using Kruskal-Wallis test and Friedman test in Matlab. The p-values obtained from the tests showed that the precipitation series from the different regions under investigation are homogeneous at 95% confidence levels. A preliminary correlation analysis was performed to get a general idea of the relationship between the precipitation and different climate indices. Table 1 shows the result of correlation analyses. The results suggest that most of the residual precipitation time series have non-significant (at 95% confidence level) correlations with the climate indices. The correlation coefficients reveal that correlation analysis is incapable of revealing the dependence between the precipitation and climate indices. However, these findings are based on the assumption of stationarity, and therefore a more appropriate method such as wavelet coherence is used in the following section to get a better understanding of the relationship.

Table 2 shows the interdependence of the climate indices for the different seasons of the year. It can be observed that some of the climatic factors are significantly (at 95% confidence levels) interdependent on each other. During the southwest monsoon months (June, July, Aug, and Sep) the relationship between Niño 3.4 and SOI, PDO are found to be statistically significant. The correlation between Nino 3.4 and SOI is clear from the fact that they represent the same ENSO phenomena. The reason for choosing both these indices in the present roots from the statement from Barnston (2015), who argues that ENSO is multifaceted, involving different aspects of the ocean and the atmosphere over the tropical Pacific. Further, one cannot measure one aspect of the entire tropical Pacific perfectly, so we get a better picture when we consider a few related measures.

Correlation analysis (table 1) also shows that the Niño 3.4 and IOD are not significantly correlated on the other hand IOD and SOI have significant correlation for some seasons. Surprisingly, the climate indices from the North Atlantic Ocean such as AMO and NAO are not correlated with the other indices during any of the seasons.

A simple wavelet coherence analysis between one climate index and precipitation will certainly carry the effect of other climate indices due to the interdependence. Therefore, it becomes imperative to independently analyse the effect of one climate index on the precipitation after removing the effect of other climate indices. Since the focus of the study is to analyse the interdependent climate indices and then study the standalone effect on precipitation by each of the climate indices after removing the other's effect we have therefore only taken those indices (Niño 3.4, SOI, IOD, and PDO) which share some level of interdependence with each other.

3. Methods

*Wavelet analysis*

Wavelet method is a multi-resolution analysis used to obtain time-frequency representations of a continuous signal. Wavelet analysis transforms a signal into scaled and translated versions of an original (mother) wavelet, instead of decomposing a signal into constituent harmonic functions as in Fourier analysis. The wavelet transform as defined by Eq. (1) (Daubechies, 1992) is called the continuous wavelet transform (CWT) because of the scale and time parameters, *a* and *τ*, assume continuous values. (Agarwal et al. (2016 a,b); Giri et al. (2014)).

$$W(a,\tau) = \frac{1}{\sqrt{|a|}} \int_{-\infty}^{\infty} f(t)\psi\left(\frac{t-\tau}{a}\right) dt \qquad (1)$$

It provides a redundant representation of a signal as CWT of a function $f(t)$ at scale '*a*' and location '*τ*' can be obtained from the continuous wavelet transform of the same function at other scales and locations. Here, $\psi$ represents the family of function called wavelets and $t$ represents the time. Since the CWT behaves like an orthonormal basis decomposition, it can be shown that it is also isometric as it preserves the overall energy content of the signal and, thereby, allows for the recovery of the function *f(t)* from its transform by using the following reconstruction formula as provided by Daubechies (1992) in Eq.(2)

$$f(t) = \frac{1}{C_\psi} \int_{-\infty}^{\infty}\int_{0}^{\infty} a^{-2} W(a,\tau) \psi_{a,\tau}(t) \, da \, d\tau \qquad (2)$$

where $C_\psi$ is a constant and depends on the choice of the wavelet $\psi$. Clearly, the above equation suggests that the function *f(t)* may be seen as a superposition of signals at different scales and obtained by varying the scale parameter '*a*'.

Further, the energy of the signal *f(t)* can be represented scale wise as given by Daubechies (1992) in Eq.(3)

$$\int_{-\infty}^{\infty} f^2(t)dt = \frac{1}{C_\psi} \int_0^\infty \left[ \int_{-\infty}^{\infty} |W(a,\tau)|^2 d\tau \right] \frac{da}{a^2} \qquad (3)$$

The left-hand side of Eq. (3) is called the 'energy' of the signal *f(t)* (it is, however, not energy in the physical sense unless *f(t)* has the proper units). We can thus interpret $[W(a,\tau)]^2 d\tau$ as being proportional to an energy density function that decomposes the energy in *f(t)* across different scales and times. Flandrin (1988) denoted the function $|W(a,\tau)|^2$ as scalogram and for two different functions *f(t)* and *g(t)*, the product of $W_f(a,\tau)$ and $W_g(a,\tau)$ may be called a cross wavelet transform (Sehgal et al. (2014)).

*Cross-wavelet transform*

While, in general, wavelet transform provides an unfolding of the characteristics of a process in the scale-space plane, a cross wavelet transform, on the other hand, provides a similar unfolding of possible interactions of two processes, and this measure can be quite revealing about the structure of a particular process or about the interaction between different processes at different scales. The cross wavelet transform (XWT) identifies the cross wavelet power of two time series. For two given discrete time series, *X (n=1…N)* and *Y (n=1…… N)*, the XWT, $W^{XY}$ is calculated using Eq. (4)

$$W^{XY}(a) = W^X(a) \times W^{Y*}(a) \qquad (4)$$

where $W_n^X(a)$ is the CWT of time series *X* and $W^{Y*}(a)$ is the complex conjugate of $W^X(a)$, the CWT of timeseries *Y*.

The cross wavelet spectrum, although very useful in detecting the phase spectrum, can potentially lead to misleading results as it is just the product of two non-normalized wavelet spectrums (Maraun and Kurths (2004)).

*Wavelet Coherence*

The wavelet coherence (WC) avoids this problem by normalizing to the single wavelet power spectrum. Consider two-time series y and $x_1$, the wavelet coherence between these two series is given by,

$$R(y, x_1) = \frac{\varsigma[W(y, x_1)]}{\sqrt{\varsigma[W(y)]\varsigma[W(x_1)]}}.$$

$$R^2(y, x_1) = R(y, x_1) \cdot R(y, x_1)^*; \tag{5}$$

where $R(y, x_1)$ is the measure of the wavelet coherence between y and $x_1$; and $R^2(y, x_1)$ is the measure of squared wavelet coherence between y and $x_1$; $W(y, x_1)$ denotes the denote corresponding cross-wavelet transforms and $W(.)$ denotes the wavelet transform; $\varsigma$ denotes a smoothing operator that can be used to balance between desired time-frequency resolution and statistical significance. The WC ranges from 0 to 1 and measures the cross-correlation of two timeseries as a function of frequency (Torrence and Compo (1997)), i.e. local correlation between the timeseries in time-frequency space. It can be interpreted as a decomposition of correlation coefficient at a different scale (Casagrande et al. (2015)); the closer the value to 1, more the correlation between the two series. Statistically, significant wavelet coherences were identified using significance test based on Grinstead et al. (2004). A total of 1,000 realizations with the same first-order autoregressive (AR1) process coefficients as the two input data sets are generated using Monte Carlo techniques. The wavelet coherence is then calculated for each of these realizations and the significance level is calculated for each scale. To understand the multiscale dependence of the precipitation on the climate indices, the wavelet coherence plots were used. The wavelet coherence was done using the Grinstead Toolbox in MATLAB.

*Partial Wavelet Coherence (PWC)*

PWC is a technique similar to the partial correlation that helps to find the resulting WC between two-time series y and $x_1$ after eliminating the influence of the timeseries $x_2$. Mihanovic et al. (2009) extended the concept from simple linear correlation and suggested that the PWC squared (after the removal of the effect of $x_2$) can be defined by an equation similar to the partial correlation squared, as shown in Eq. (6) which is like the simple WC, ranging from 0 to 1.

$$RP^2(y, x_1, x_2) = \frac{\left| R(y, x_1) - R(y, x_2).R(y, x_1)^* \right|^2}{\left[ 1 - R(y, x_2) \right]^2 \left[ 1 - R(x_2, x_1) \right]^2} \quad (6)$$

*where $RP^2$ denotes the squared partial wavelet coherence.*

R (.,.) denotes the wavelet coherence between the two variables and $RP^2$(y, x1, x2) is the partial wavelet coherence squared between *y* and $x_2$ when the influence of $x_1$ is excluded. Its proximity to zero at a certain time-frequency point indicates that the series $x_2$ does not add significant information to y, i.e. the information that is not already incorporated from $x_1$ at that point. If partial wavelet coherence squared is high for $x_1$ and not for $x_2$, this would imply that important covariance exists between $x_1$ and *y* during that time interval at a designated wavelet scale (period), and moreover that *y* was dominantly influenced by $x_1$ and not by $x_2$. If both $RP^2$(y, $x_1$, $x_2$) and $RP^2$(y, $x_2$, $x_1$) still have significant bands, both $x_1$ and $x_2$ have a significant influence on y. In this study, the toolbox provided by Ng and Chan (2012) is used for estimating the partial wavelet coherence.

## 4. Results

### 4.1 Wavelet Coherence analyses

To understand the variability of precipitation with reference to time, Continuous Wavelet Transform (CWT) was performed on the series using 'morlet' wavelet. Morlet wavelet was chosen because it is widely used complex wavelet having good time-frequency localization than other real wavelets (Addison (2002)). However, it is to note that Kumar and Foufoula (1994) and Gan et al. (2007) have shown that in this kind of multiscale coherence analysis the choice of wavelets does not make much difference. The wavelet coefficients obtained from CWT for the precipitation has been plotted (not shown here) and the significance levels are calculated according to Grinstead et al. (2004). Using the CWT plot, the dominant periods and their variability in terms of time were identified. The summary of the results is tabulated in Table 3. It can be seen that in most of the stations, the long term oscillation exist having a periodicity of 64-128 months (~5 to 10 years) or even greater in some cases.

To understand the underlying causative factors driving the low-frequency precipitation process, wavelet coherence analysis was performed between the observed precipitation and climate indices.

*Relationship between* Niño 3.4 *and precipitation*

Even though all the analysis was done for 30 stations, however, results are shown only for 15 stations for brevity. Figure 2 shows the wavelet coherence plot between the standardized precipitation series and Niño 3.4 for the different stations across India. It is found that for all the regions there is a very high correlation between Niño 3.4 and precipitation anomalies at the scale of 2-4 years, 4-7 years and 8-16 years. There was no significant effect of Niño 3.4 on the interannual dynamics of the Indian precipitation. It is observed that along with high coherence time intervals, intervals of weakening coupling or even the absence of significant coupling between Niño 3.4 and precipitation are detected. The analysis of the plot reveals that the coupling between Niño 3.4 and precipitation has become stronger in the recent decades (as shown by the red areas after 1960) in most parts of the country. Further, closer analyses (as shown for one station) of the directions of the arrows indicate that the relationship between the ENSO and precipitation is bidirectional coupling with the alteration of driving and driven process. Also, there is a significant difference in the wavelet coherence plots for different regions. Precipitation at stations such as Madurai, Bangalore, Varanasi, and Delhi

Jaipur does not have significant coherence with Niño 3.4 at 16-32 years scale, whereas other stations such as Chennai, Kolkata, Gandhinagar, Bandhra, Shillong have significant Niño 3.4 influence at that large scale.

*Relationship between PDO, SOI, and precipitation*

Figures 3 and 4 show the wavelet coherence analysis between PDO, SOI, and the precipitation anomalies, respectively. Notably, the multiscale coherence plot reveals a pattern similar to Nino 3.4 that exists between PDO, SOI, and precipitation in most of the stations. It can be observed that majority of the features present in Figure 2 exist in the coherence plots between PDO, SOI, and precipitation. However, it is to be noted that there is a clear observation of a reduction of power (a measure of the degree of coherence) when compared to the ones obtained for Nino 3.4.

*Relationship between IOD and precipitation*

Figure 5 shows the wavelet coherence plot between IOD and precipitation. It can be seen that for most of the stations, there are consistent significant long-term features having periods greater than 8 years and 16 years indicating a high correlation between IOD and precipitation at these scales. Apart from this consistent long-term feature, there are only a few intermittent high power regions. The above analysis revealed the relationship between the climate indices and Indian precipitation is spatially and temporally variable. It can be observed that from wavelet coherence analysis, the precipitation appears to be related to all the factors Niño 3.4, IOD, PDO, and SOI but at different scales and at varying degrees. However, it should be noted that there was a significant level of interdependence across these climate indices as shown in Table 2. Therefore, the above coherence plots of the precipitation with individual indices might be misleading as there might be the carry over effect of one index over the other. And for a better understanding of the dynamics of the influence of the climate indices, it is necessary to understand the stand alone effect of the indices on the precipitation. For this purpose, we have used partial wavelet analysis to study the standalone effect of the climate indices on precipitation.

In the past, studies such as Newman et al. (2003) showed the dependence of PDO on ENSO events (Niño 3.4 or SOI) and Ashok et al. (2004) have discussed the relationship between IOD and ENSO events (Niño 3.4 or SOI). From these studies, it is clear that ENSO events (Niño 3.4 or SOI) play an influencing role in controlling the other climate indices. Therefore, the coherence between the standardized precipitation and each of the climate indices (PDO, SOI, and IOD) after removing the effect of Niño 3.4 is examined. This would help in understanding the standalone effect of the SOI, IOD, and PDO on Indian precipitation.

**4.2 Partial wavelet Coherence Analysis**

Figures 6, 7, 8 show partial wavelet coherence between the standardized precipitation and climate indices (PDO, SOI, and IOD) after removing the effect of Niño 3.4. Figure 6 shows the wavelet coherence analysis between precipitation and PDO after removing the effect of Niño 3.4. When compared with Figure 3 (which shows the wavelet coherence analysis between PDO and precipitation) a significant reduction in the high power regions is observed. Also, the correlation between the precipitation and PDO is found to be zero in most of the stations. For example, in the case of Chennai, Figure 3 showed a significant relationship between precipitation and PDO at 4-8 years scale. However, such trend is not observed in Figure 6.

Similarly, Figure 7 shows the standalone relationship between precipitation and SOI after removing the influence of Niño 3.4. Here, again, a significant difference between Figure 7 and Figure 4 (which shows the wavelet coherence analysis between SOI and precipitation) is revealed. It is observed that there is not a strong linkage between the SOI and Indian precipitation except at few places where there is a good degree of correlation at 10 years scale. Apart from few pockets of high energy regions (in Chennai, Madurai, Kodugu), there is no significant relationship between the SOI and precipitation.

On the contrary, there is barely a change in the wavelet coherence plots between IOD and precipitation before and after removing the effect of Niño 3.4. Figure 8 shows the wavelet coherence plot between IOD and precipitation after removing the effect of Niño 3.4. There exists a significant relationship between precipitation and IOD scales greater than 10 years. This indicates that the IOD is not influenced by the Niño 3.4and has independent influence over the precipitation in most of the regions in India.

The results reveal that the in most part of India, precipitation is influenced by Niño 3.4and IOD but at different time scales. Therefore, the subsequent analysis considers only Niño 3.4and IOD as potential factors of variability in Indian precipitation.

**4. 3 Reconstruction of oscillatory modes**

To further understand the drivers of precipitation variability, modes at significant frequencies were extracted from precipitation time series and correlated with oscillations associated with sea-surface temperature and atmospheric forcing, specifically Niño 3.4and IOD. For this purpose, the wavelet coefficients were reconstructed at different bands of frequencies. Figure 9a shows the scale (frequency)-wise component of IOD and precipitation for one sample station (Chennai) using db5 wavelets for seven levels of decomposition. Similarly, Figure 9b shows the scale-wise component of Niño 3.4and precipitation at Chennai for seven levels of decomposition. Similar analysis was done for other stations and the results were summarized in Table 4 in the form of correlation coefficients. The analyses of the correlation analysis at multiple scales reveal that the association between precipitation modes and the climate indices oscillatory modes is spatially variable. It was observed that a significant association was found only at long-term scales of the order of 64-128 months with IOD and at comparatively short-term scales of the order of 16-32 and 32-64 months scales with Niño 3.4. It was also seen that the association precipitation - IOD and precipitation- Niño 3.4was complementary to each other. In most of the places, it was observed that when the association with IOD was strong the association with Niño 3.4was weak and vice-versa. However, at some of the places like Varanasi, Jabalpur Aligarh, Puri, Jalgaon, and Amravati the association was significant with both IOD and ENSO.

Figure 10 shows the box plot of the correlation coefficients between Niño 3.4and precipitation reconstructed modes at different scales. Figure 11 shows the similar plot between IOD and precipitation. It can be seen that Niño 3.4 has a significant negative correlation at the scale of 32-64 months (2-5 years) whereas the IOD has a high positive correlation in the 64 months (10 years) scale. Figure 12 and 13 shows the season-wise split of the relationship between Niño 3.4, IOD, and precipitation. The analysis of the box plots reveals that the Niño 3.4affects the precipitation at the scales of 2- 5 years in almost all the seasons. Particularly during the summer monsoon, the effect of Niño 3.4is found to be significant at 2 months scales. On the other hand, IOD influences the precipitation at 10 years scale in all the seasons in a consistent manner. However, it is to be noted that the above analysis provides only an overall estimate of the relationship of precipitation with Niño 3.4and IOD. Understanding the fact that the relationship evolves with time due to other physical factors such as solar activity (Narasimha and Bhattacharya (2009)), we attempted to analyse the temporal evolution of the dependence between Niño 3.4, IOD, and precipitation.

**4.4 Temporal evolution of dependence between ENSO, IOD, and precipitation.**

The wavelet coherence analysis revealed that the relationship between Niño 3.4, IOD and precipitation is intermittent and varying with time. Therefore, to further understand the time-evolution of the relationship between the precipitation and the climate indices at different scales, the entire time span was split into 4 segments of 25 years each starting with 1901.

Figure 14 shows the evolution of the correlation coefficient between Niño 3.4and precipitation modes with time. Figure 15 shows the correlation coefficient between IOD and precipitation modes for different time periods. It can be seen that the relationship between the Niño 3.4and precipitations is clearly evolving with time at all the scales. During the period 1901-1925, it was observed that Niño 3.4-monsoon connection is strong (but negative), while the IOD-monsoon connection is insignificant; during 1926–1950 where both the Niño 3.4and IOD influences were weak and insignificant. For the period of 1951-1975, both the connections strongly influenced the precipitation. And for 1975-2000, the connection was relatively weak except for the decadal cycles.

## 5. Discussion

The above study is the first to analyse the stand alone effect of various climatic factors on precipitation after removing the other potential drivers. Firstly, the basic wavelet coherence analysis among different climate indicators and Indian precipitation revealed that there is strong relationship existed between precipitation and all the climatic indices analysed in this study. However, subsequent analysis using partial wavelet coherence analysis uncovered some interesting facts about the relationship amongst precipitation and climatic factors. The PWC analysis showed that effect of PDO and SOI on precipitation is just an artifact of their dependence on Niño 3.4. Also, it was observed that Niño 3.4 is the primary driver of the variability of precipitation in most of the stations at scales of the order of 2-4 and 4-7 years (Narasimha and Bhattacharyya (2009)). It was observed that IOD on the other hand, an independent driver, acting at the scale of 16 years and complements the Niño 3.4. Further, the analysis of the direction of the arrows in the coherence plot between Precipitation and Niño 3.4 showed that the Niño 3.4 and precipitation are bidirectional coupling process with the alteration of driving and driven process. This is in congruence with the observations by Mokhov et al. (2011) where the authors show that the Niño 3.4-Indian precipitation has a non-symmetric bidirectional coupling using the causality analysis.

Additionally, it can be noticed that partial wavelet analysis can be used as a useful technique in comprehending the standalone impacts of climatic indices and precipitation and in other geophysical processes. Further, the IOD events are appeared to happen independently of Niño 3.4 events as the PWC amongst IOD and precipitation is unaffected after removing the Niño 3.4 effect. Likewise, it was clear from the investigation that the scale at which the IOD and ENSO influence the precipitation are distinctive. Notwithstanding, the strong correlation amongst IOD and Niño 3.4 amid the autumn as appeared in Table 2 might be because of the co-occurrence of the Niño 3.4 and the IOD event (Ashok et al. (2004)) which fundamentally may not be because of the dependence on each other.

We also observed that the coupling between precipitation and Niño 3.4 is becoming stronger in the latter half of the analysed time (particularly after 1970). Cherchi and Navarra (2013) obtained similar results using a different approach and explains that this kind of behavior may be due to the changes occurred in the North Pacific after 1976 (Miller et al. (1994)) and to the associated differences in the Niño 3.4 teleconnections (Deser and Blackmon (1995)).

With reference to the spatial variation of the association of precipitation with Niño 3.4 and IOD, it was observed that most parts of India except central India there exists a complementary relationship of precipitation with Niño 3.4 and IOD. However, it was seen in some parts of central India (including the east and west parts) the Niño 3.4 and IOD were simultaneously influencing the precipitation but at different temporal scales. Further, it was observed that the coastal regions of peninsular and northeast and west regions were more significantly affected by Niño 3.4 and IOD in comparison with their counterparts in the inland region. The strength of the relationship between precipitation and climate indices for given period might be driving the degree of spatial variability in Indian precipitation as observed by Ghosh et al. (2012).

Overall, different climatic indices were analysed and the results show that the effect of PDO and SOI on precipitation is insignificant, contrary to the results obtained by Krishnamurthy and Krishnamurthy (2013) where the authors observe a statistically significant influence of PDO on Indian precipitation. This difference in the results warrants future scope for understanding the complex dynamics of PDO (Wang et al. (2014)).

The study also reveals the relationship between ENSO, IOD and the monsoon is not constant in time. In fact, in the record analysed (i.e. 1901-2002) it was possible to distinguish periods with the strong Niño 3.4-monsoon connection, lack of IOD-monsoon connection and increasing Niño 3.4-IOD relationship, or with both strong and negative IOD-monsoon and Niño 3.4-monsoon connections, or with not-significant monsoon- Niño 3.4 and monsoon-IOD but with strong Niño 3.4-IOD relationships. This temporal variability in the effect of the influence of IOD and Niño 3.4 on monsoon can be attributed to the changes in the North Pacific and in the Atlantic sectors.

## 6. Conclusions

Thirty precipitation time series from different parts of India were examined with the target of investigating the impacts of climatic factors on Indian precipitation. The conclusion of the study are as follows:

1) Simple correlation analysis did not demonstrate any evidence of a relationship between the climate factors and the precipitation. However, multiscale analyses using wavelets suggest that the precipitation shows a significant relationship with climatic factors such as Niño 3.4, SOI, PDO, and IOD. Further, the partial wavelet analysis revealed that SOI and PDO do not significantly affect precipitation, but seems the other way because of their interdependence on Niño 3.4. It was observed that partial wavelet analysis well-suited to bringing out the standalone relationship of climatic factors with precipitation after removing the other potential factors.

2) The wavelet coherence plot and the reconstruction of modes at different scales reveal the Niño 3.4 effects i.e. there is a strong relationship between the Niño 3.4 & precipitation at 2-7 years' scale also IOD &precipitation shows high coherency at the decadal or higher scale.

3) It was observed that the relationship between IOD, Niño 3.4, and precipitation varies with time and space.

Therefore, we conclude that the Niño 3.4 and IOD indices show significant linkages with high- and low-frequency variability in the precipitation, respectively. The results of spectral analyses coincide with the conclusions of other previous studies, which have detected the influence of Niño 3.4 and IOD on hydrologic variables in India.

The present study only investigated the linear relationship between the climate indices and the precipitation. However, it will be interesting to further investigate the presence of nonlinear dependence between them. Correlation does not imply causation, therefore, another scope for further study is to investigate the causal relationship between the precipitation and climate indices at different scales and hence use such information for enhanced precipitation forecast.

**Acknowledgements**

This research was funded by Department Science and Technology, India through the INSPIRE Faculty Fellowship held by Maheswaran Rathinasamy. We are grateful for the two anonymous reviewers for their valuable suggestion in improving the presentation and manuscript.

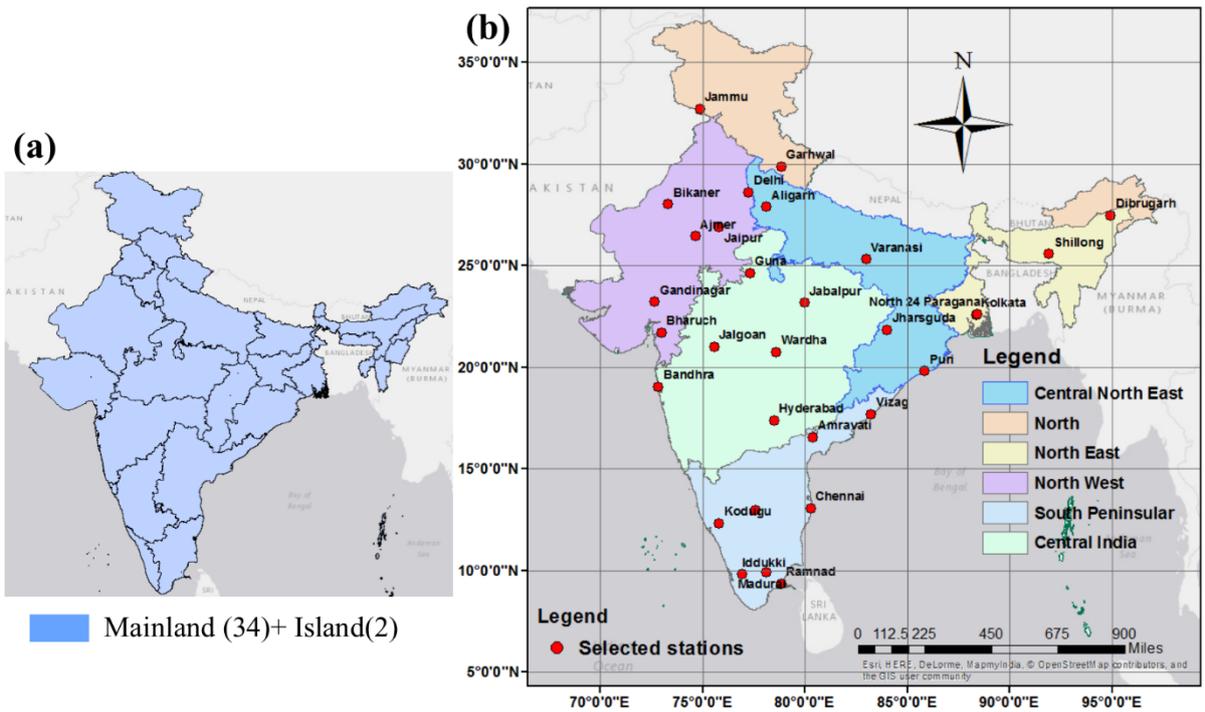

**Figure 1 (a) Indian terrain showing 34 mainland and 2 Island (b) Geographic location of stations under investigation**

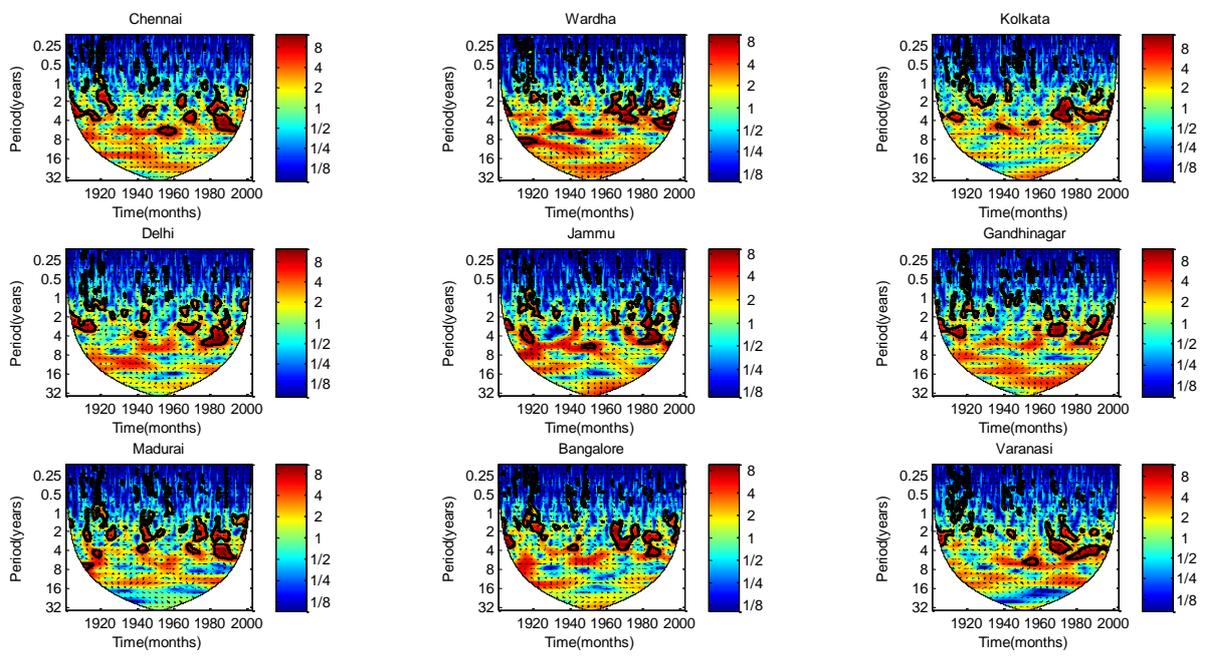

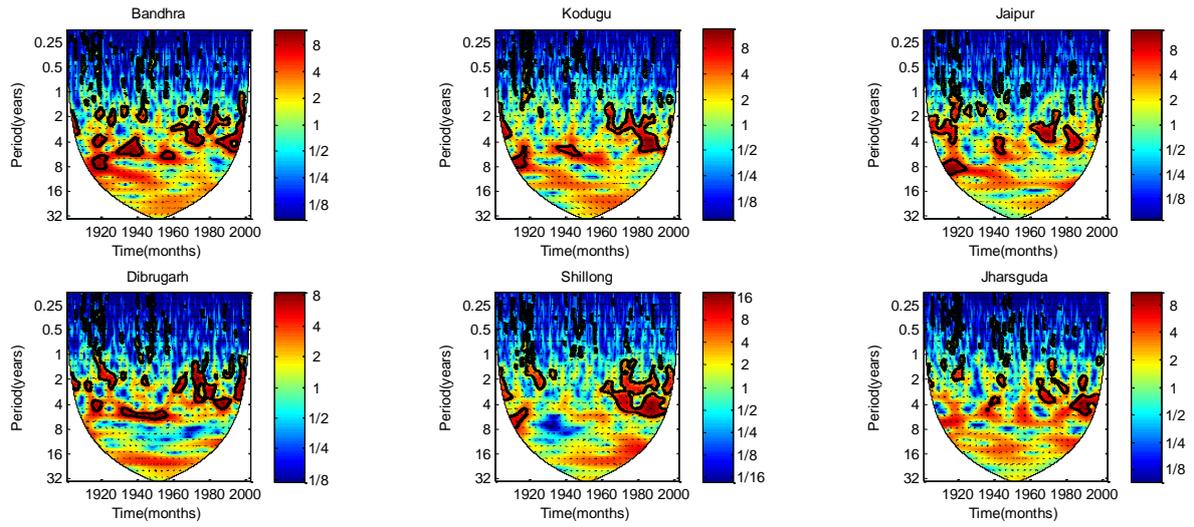

**Figure 2: Plot showing wavelet coherence between the standardized precipitation and NIÑO 3.4 at different stations. The sections marked by black color contours are the significant sections indicating high coherence between precipitation and NIÑO 3.4. The cone of influence indicates the area affected by the boundary assumptions.**

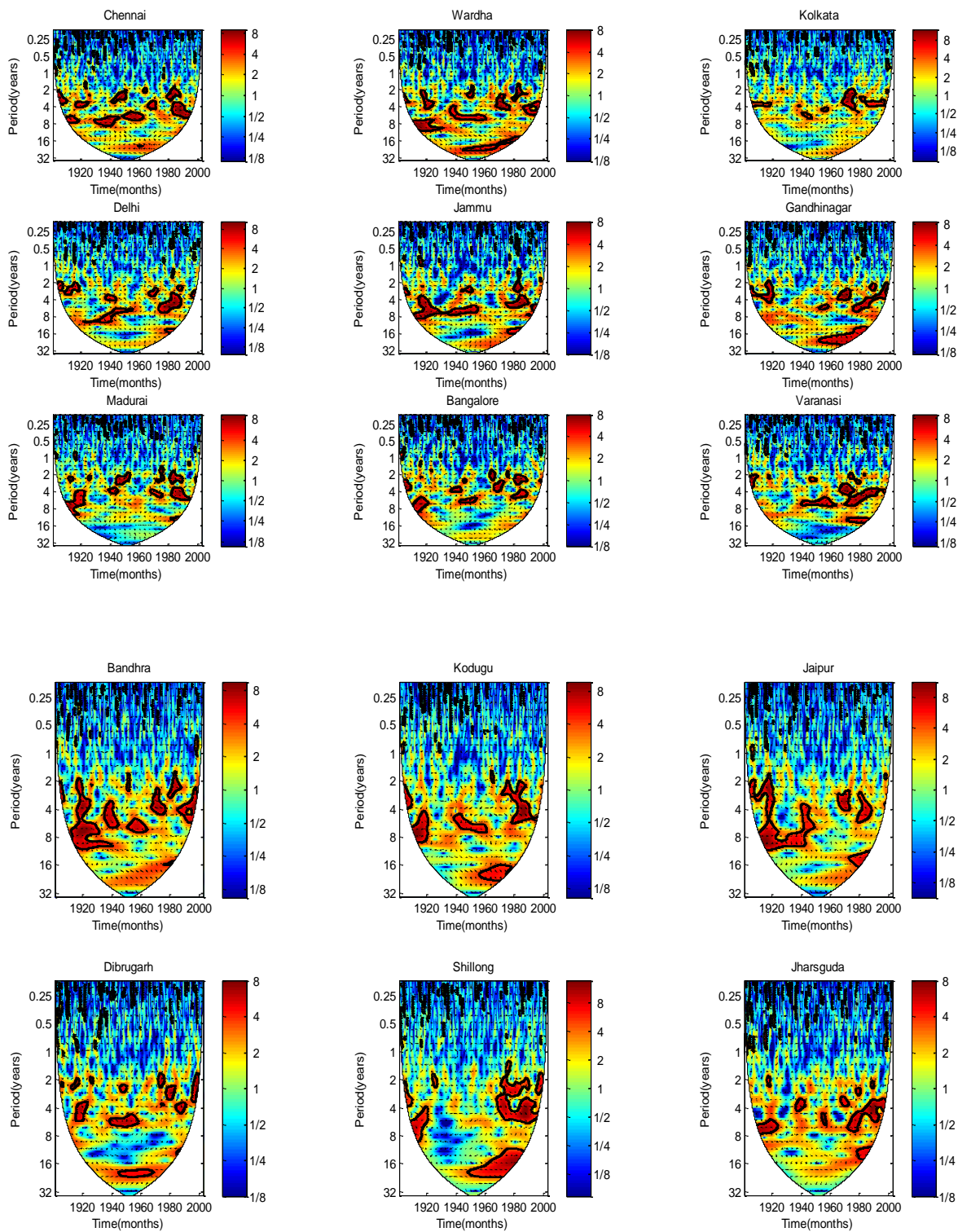

**Figure 3:** Plot showing wavelet coherence between the standardized precipitation and PDO at different stations. The sections marked by black color contours are the significant sections indicating high coherence between precipitation and PDO. The cone of influence indicates the area affected by the boundary assumptions.

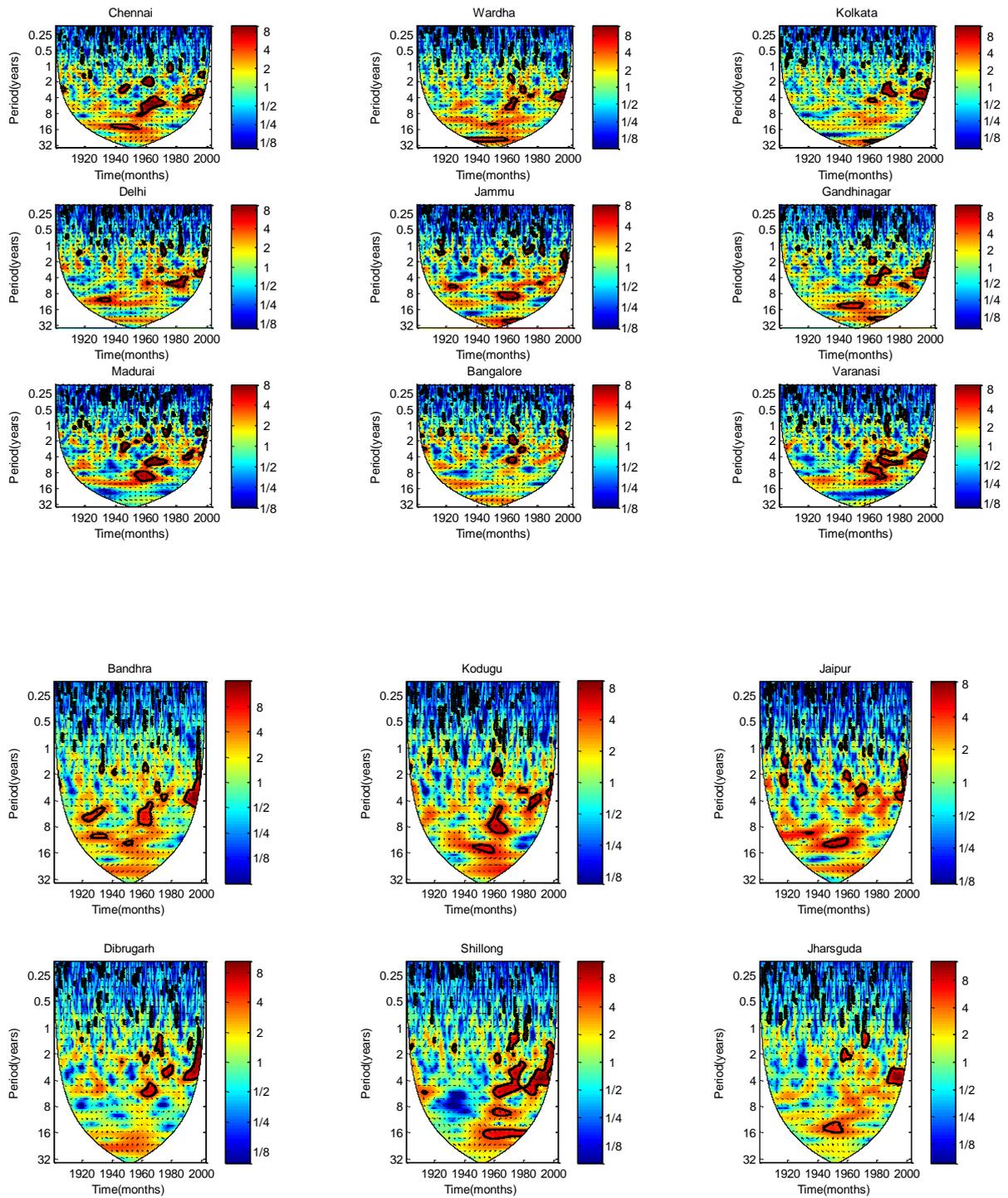

**Figure 4: Plot showing wavelet coherence between the standardized precipitation and SOI at different stations. The sections marked by black color contours are the significant sections indicating high coherence between precipitation and SOI. The cone of influence indicates the area affected by the boundary assumptions.**

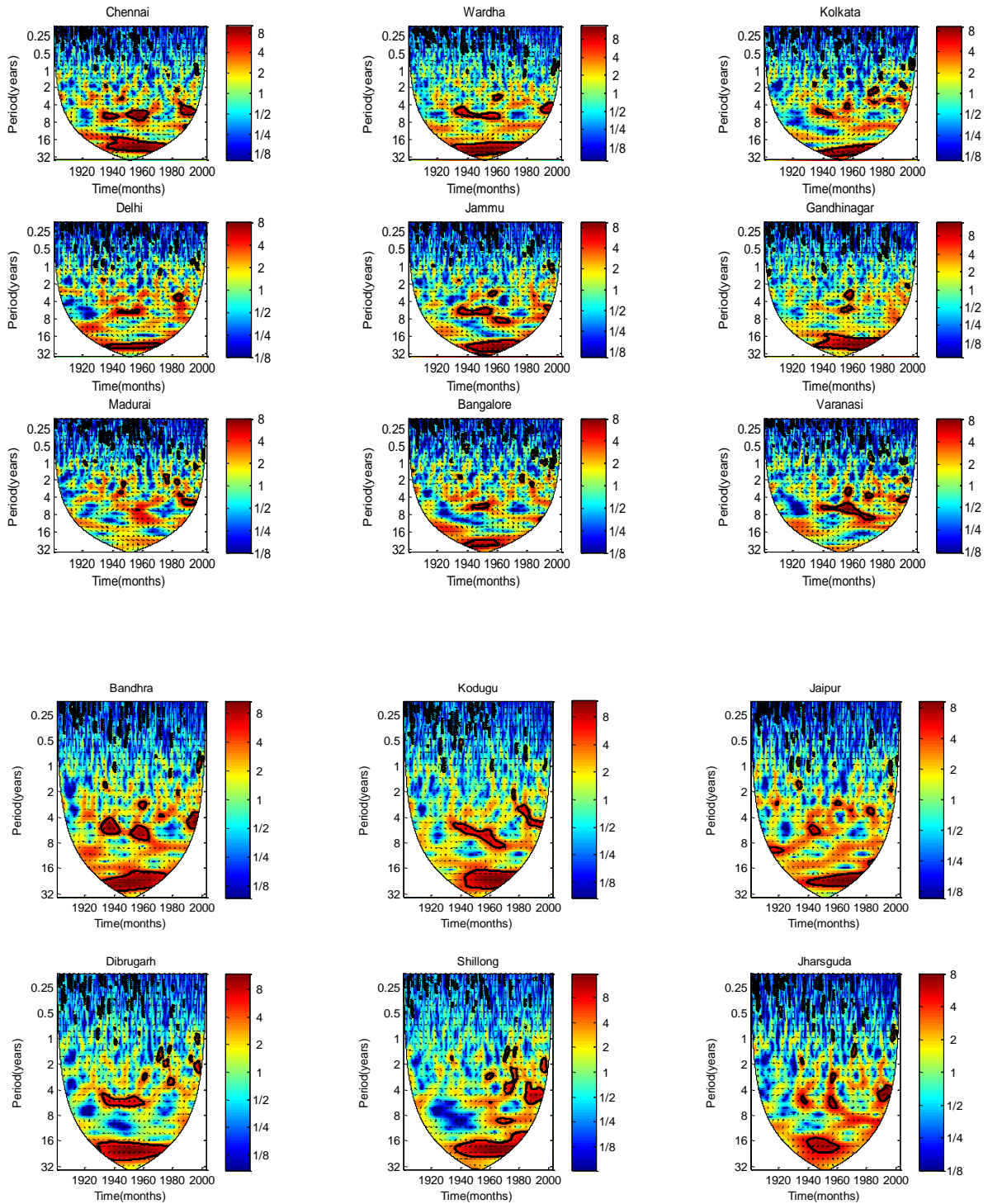

**Figure 5:** Plot showing wavelet coherence between the standardized precipitation and IOD at different stations. The sections marked by black color contours are the significant sections indicating high coherence between precipitation and IOD. The cone of influence indicates the area affected by the boundary assumptions.

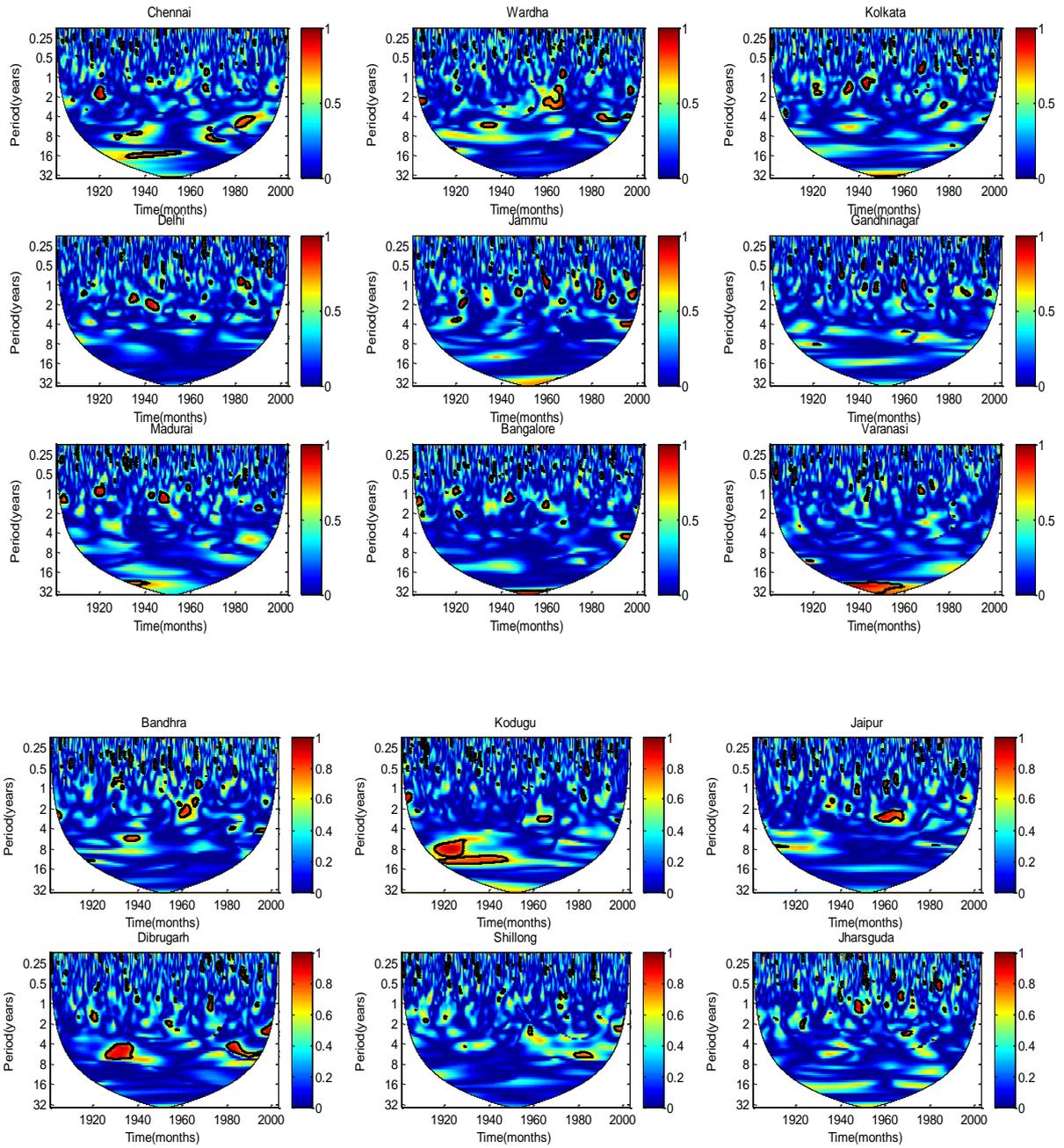

**Figure 6 Plot showing partial wavelet coherence between the standardized precipitation at different stations and PDO after removing the effect of NIÑO 3.4. The sections marked by red color indicate high correlation between the two variables.**

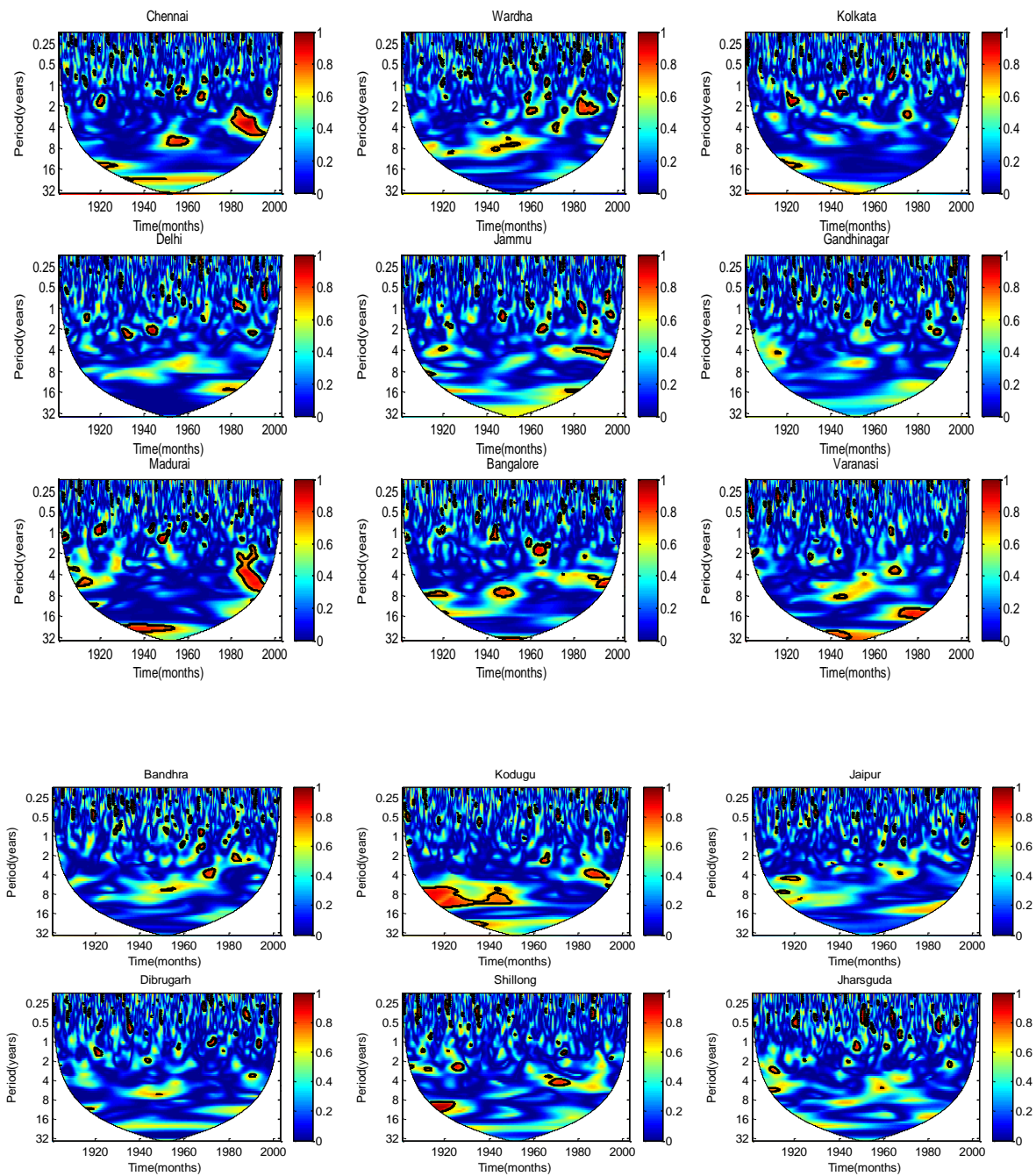

**Figure 7 Plot showing partial wavelet coherence between the standardized precipitation at different stations and SOI after removing the effect of NIÑO 3.4. The sections marked by red color indicate high correlation between the two variables**

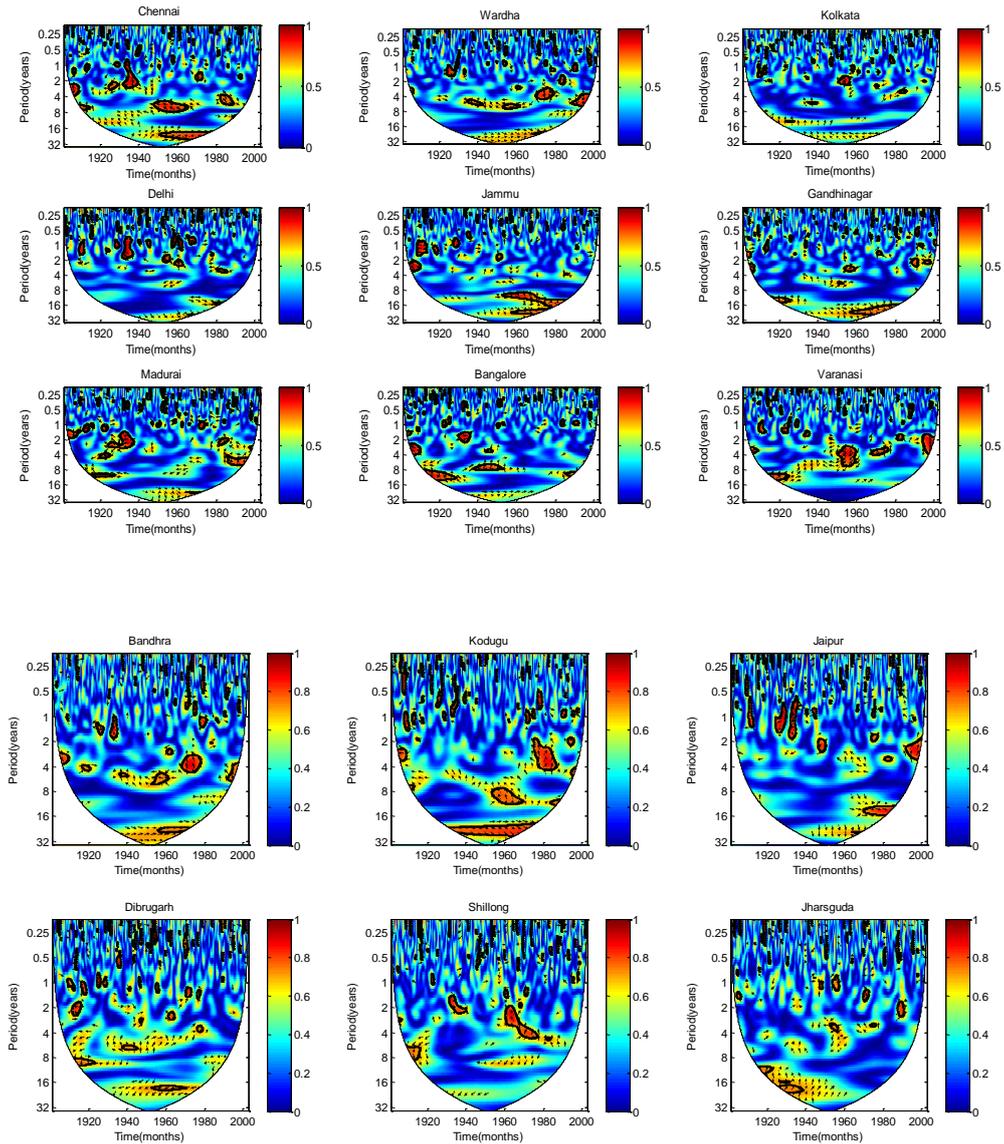

**Figure 8 Plot showing partial wavelet coherence between the standardized precipitation at different stations and IOD after removing the effect of NIÑO 3.4. The sections marked by red color indicate high correlation between the two variables**

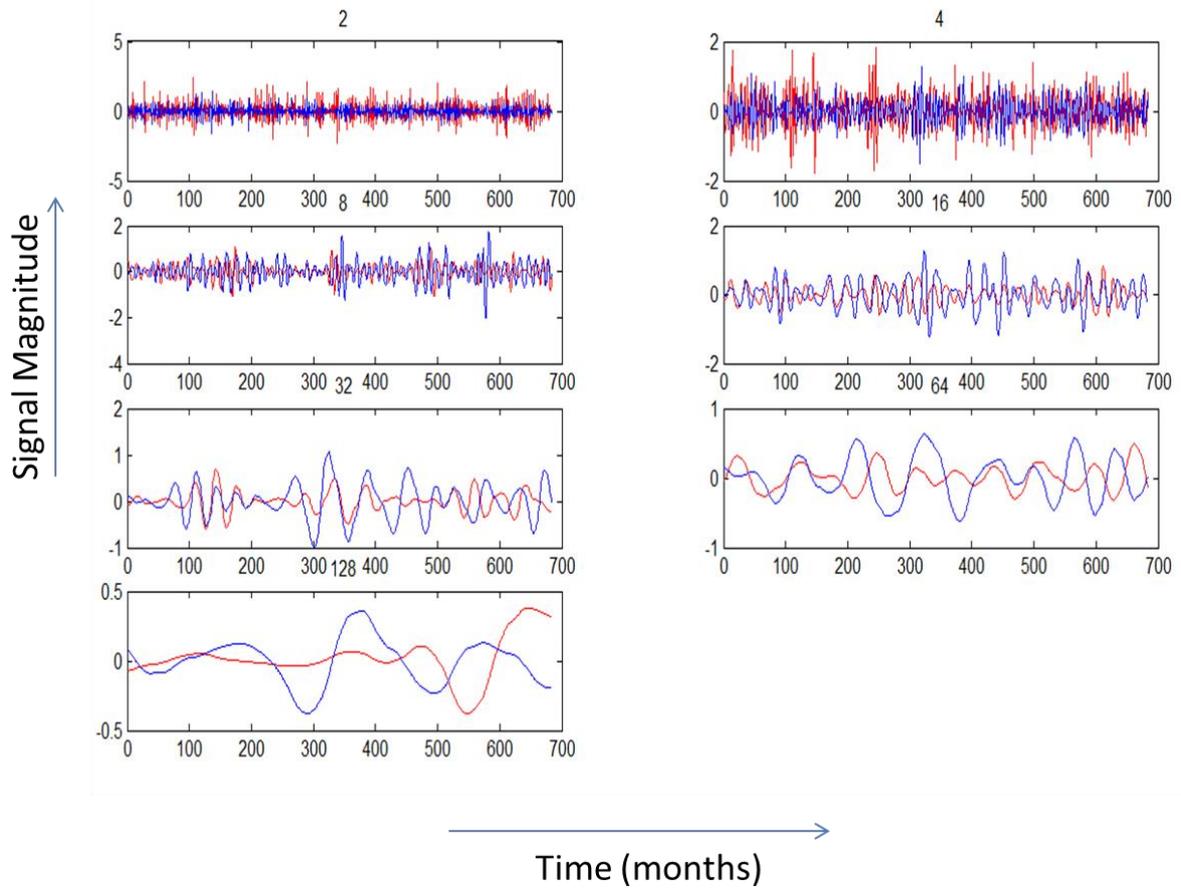

**(a)**

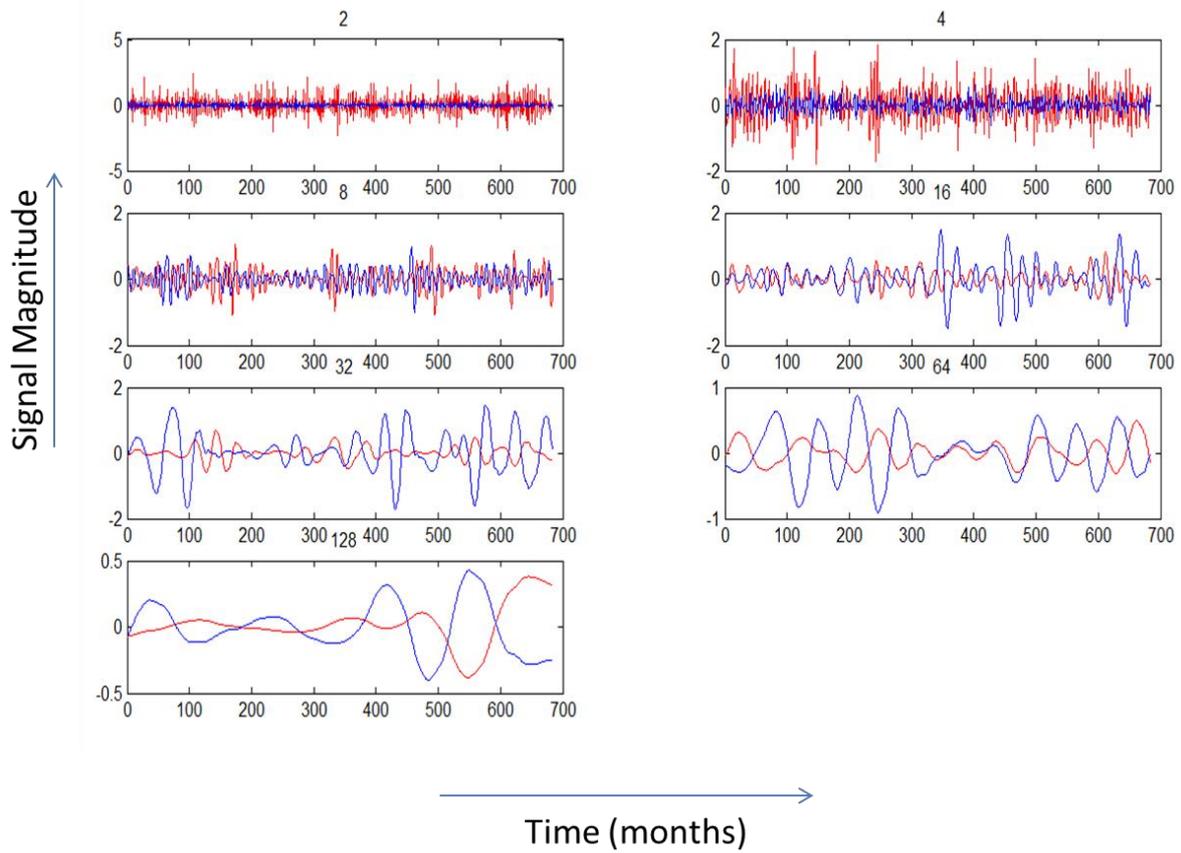

**(b)**

**Figure 9** Wavelet Decomposition of (a) IOD and (b) NIÑO 3.4along with the precipitation of Chennai (red colour) at different scales (2, 4, 8, 16, 32, 64 and 128 months). The horizontal axis denotes the time in months and vertical axis denotes the magnitude of the signal.

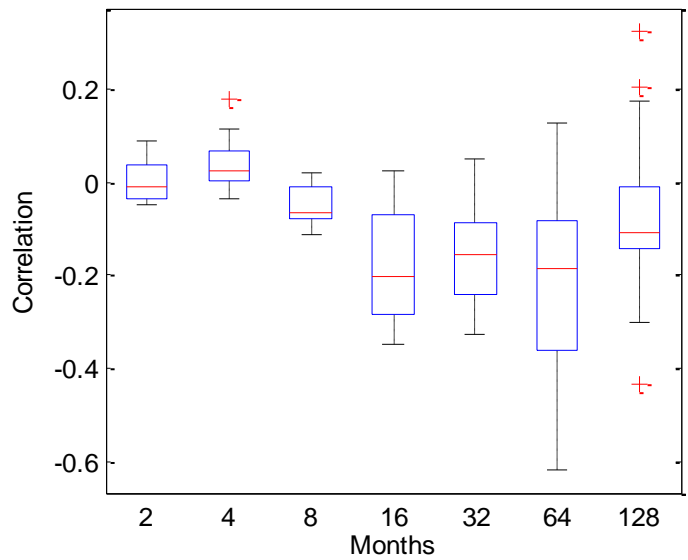

**Figure 10 Box plot showing the variation of correlation between NIÑO 3.4 and precipitation at different scale for all the stations analysed (N=30)**

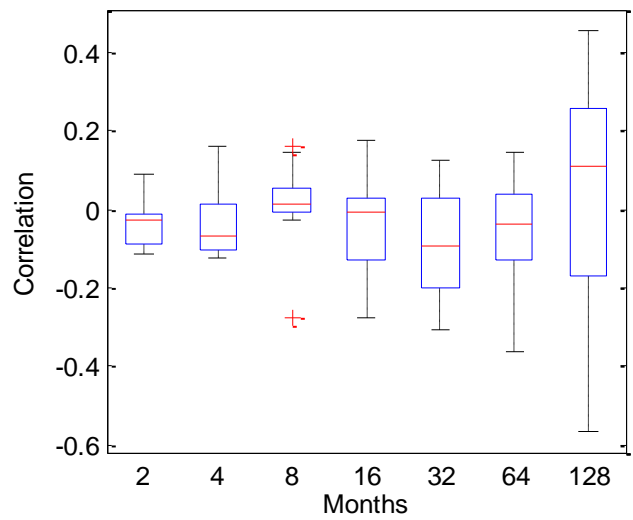

**Figure 11 Box plot showing the variation of correlation between IOD and precipitation at different scale for all the stations analysed (N=30)**

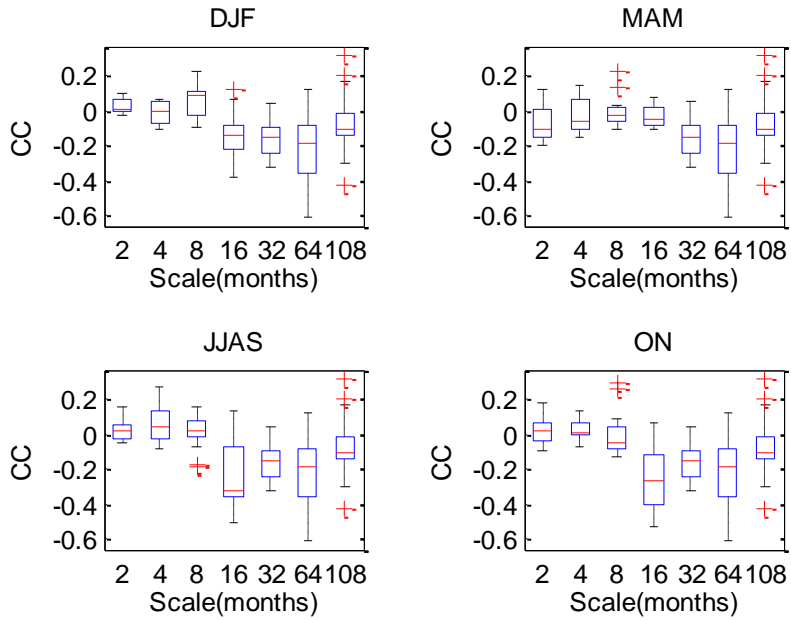

**Figure 12 Box plot showing the season-wise variation of correlation between NIÑO 3.4 and precipitation at different scales for all the stations analysed (N=30)**

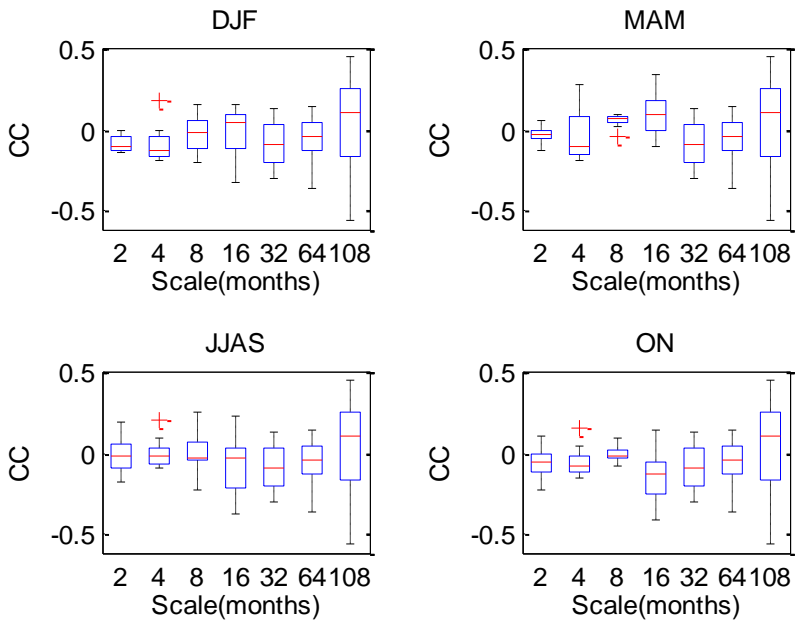

**Figure 13 Box plot showing the season-wise variation of correlation between IOD and precipitation at different scales for all the stations analysed (N=30)**

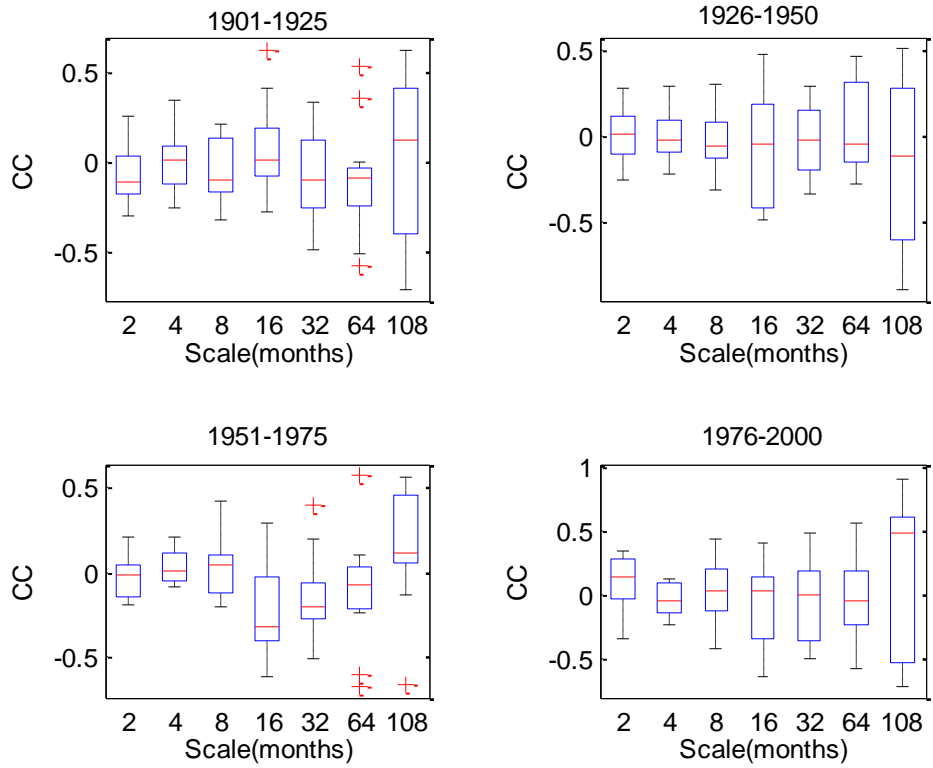

**Figure 14 Box plot showing the variation of correlation between IOD and precipitation (across 30 stations) at different scales for different time periods**

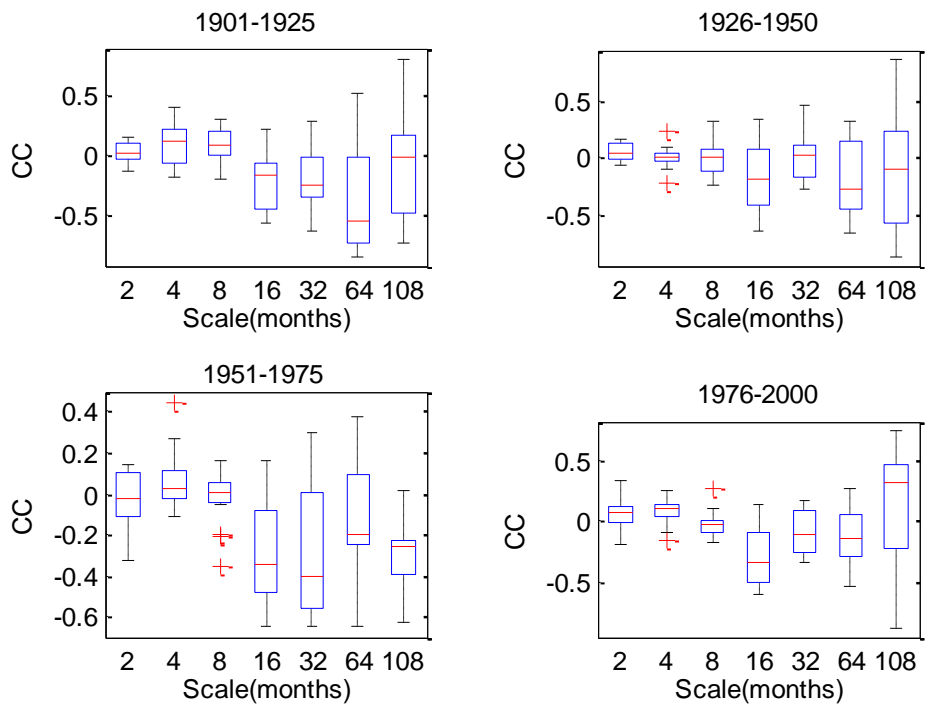

**Figure 15 Box plot showing the variation of correlation between NIÑO 3.4and precipitation (across 30 stations) at different scales for different time periods**

**Table 1. Correlation Coefficients for Monthly Standardized Precipitation at different stations and Monthly Standardized Climate Indices**[a]

| Station | Lat | Long | NINO3.4 | PDO | SOI | IOD | NAO | AMO |
|---|---|---|---|---|---|---|---|---|
| Ajmer | 26.45 | 74.64 | -0.06 | 0.02 | 0 | -0.02 | 0.04 | -0.01 |
| Aligarh | 27.90 | 78.09 | 0.02 | -0.05 | 0 | -0.01 | 0.05 | -0.01 |
| Amarvati | 16.57 | 80.36 | -0.09 | 0.04 | -0.05 | -0.02 | -0.02 | 0.07 |
| Bandhra | 19.06 | 72.84 | -0.03 | -0.03 | -0.01 | 0 | 0.03 | 0.05 |
| Bangalore | 12.97 | 77.59 | -0.04 | 0.07 | -0.05 | 0.01 | 0 | 0.05 |
| Bharauch | 21.71 | 73.00 | -0.04 | 0.05 | -0.03 | -0.04 | 0.03 | -0.03 |
| Bikaner | 28.02 | 73.31 | -0.03 | -0.01 | 0 | 0 | 0.01 | -0.05 |
| Chennai | 13.08 | 80.27 | -0.02 | 0.07 | -0.01 | 0 | 0.02 | -0.01 |
| Delhi | 28.61 | 77.21 | 0.02 | -0.05 | 0.02 | -0.02 | 0.04 | -0.01 |
| Dibrugarh | 27.47 | 94.91 | -0.03 | 0.02 | -0.04 | 0.01 | 0.05 | 0.03 |
| Gandhinagar | 23.22 | 72.64 | -0.04 | 0.01 | -0.02 | -0.06 | 0.04 | -0.01 |
| Garhwal | 29.87 | 78.84 | -0.01 | 0.01 | -0.02 | -0.02 | 0 | -0.02 |
| Guna | 24.63 | 77.30 | -0.03 | -0.01 | 0.01 | 0 | 0.02 | 0.05 |
| Hyderabad | 17.39 | 78.49 | -0.03 | 0.07 | -0.05 | -0.02 | 0.03 | 0.03 |
| Iddukki | 9.81 | 76.93 | -0.03 | 0.11 | 0.01 | -0.05 | 0.02 | 0.04 |
| Jabalpur | 23.18 | 79.99 | -0.04 | 0.01 | -0.01 | -0.04 | 0.01 | 0.05 |
| Jaipur | 26.91 | 75.79 | -0.05 | 0.01 | 0 | -0.02 | 0.05 | -0.01 |
| Jalgoan | 21.01 | 75.56 | -0.05 | 0.04 | -0.04 | 0.01 | 0.01 | 0.04 |
| Jammu | 32.72 | 74.86 | 0.06 | -0.04 | 0.02 | -0.01 | 0.08 | -0.05 |
| Jharsguda | 21.86 | 84.01 | 0.03 | -0.06 | 0 | -0.02 | 0.07 | 0.01 |
| Kodugu | 12.34 | 75.81 | -0.07 | 0.08 | -0.07 | -0.04 | 0.02 | 0.04 |
| Kolkata | 22.57 | 88.36 | -0.01 | -0.02 | 0.02 | 0 | 0.05 | -0.01 |
| Madurai | 9.93 | 78.12 | -0.04 | 0.12 | 0.01 | -0.03 | 0.02 | 0.04 |
| North24Paragana | 22.62 | 88.40 | -0.01 | -0.02 | 0.02 | 0.01 | 0.05 | -0.02 |
| Puri | 19.81 | 85.83 | 0.04 | -0.03 | -0.01 | -0.04 | 0.06 | 0.02 |
| Ramnad | 9.37 | 78.83 | -0.03 | 0.11 | 0.01 | -0.01 | 0.02 | 0.03 |
| Shillong | 25.58 | 91.89 | -0.04 | 0.05 | -0.04 | 0.01 | 0.07 | -0.07 |
| Varanasi | 25.32 | 82.97 | 0 | 0.01 | 0.02 | -0.04 | 0.04 | 0 |
| Vizag | 17.69 | 83.22 | 0.05 | -0.02 | -0.01 | 0 | 0.08 | 0.03 |
| Wardha | 20.75 | 78.60 | -0.07 | 0 | -0.05 | 0 | -0.01 | 0.06 |

[a]Bold coefficients are significant at alpha = 0.01.

**Table 2 Correlation Coefficients between the different climate indices**

| Months | | NINO3.4 | PDO | SOI | IOD | NAO | AMO |
|---|---|---|---|---|---|---|---|
| DJF | NINO3.4 | 1 | | | | | |
| | **PDO** | **0.39203** | 1 | | | | |
| | **SOI** | **0.77582** | **0.42541** | 1 | | | |
| | IOD | 0.127694 | 0.006052 | 0.149418 | 1 | | |
| | NAO | -0.0014 | -0.001 | 0.001 | -0.0053 | 1 | |
| | AMO | 0.045 | -0.025 | 0.172 | -0.005 | -0.116 | 1.000 |
| MAM | NINO3.4 | 1 | | | | | |
| | **PDO** | **0.382566** | 1 | | | | |
| | **SOI** | **0.646109** | **0.24767** | 1 | | | |
| | IOD | 0.129083 | 0.072002 | **0.233378** | 1 | | |
| | NAO | 0.025 | -0.090 | 0.015 | -0.049 | 1 | |
| | AMO | 0.266 | -0.121 | 0.002 | 0.169 | -0.100 | 1 |
| JJAS | NINO3.4 | 1 | | | | | |
| | **PDO** | **0.500945** | 1 | | | | |
| | **SOI** | **0.767654** | **0.33789** | 1 | | | |
| | **IOD** | **0.15391** | 0.0693 | **0.193718** | 1 | | |
| | NAO | -0.0148 | -0.048 | 0.072 | -0.054 | 1 | -0.125 |
| | AMO | 0.002 | 0.028 | -0.087 | 0.077 | -0.125 | 1 |
| ON | NINO3.4 | 1 | | | | | |
| | **PDO** | **0.375404** | 1 | | | | |
| | **SOI** | **0.765836** | **0.318892** | 1 | | | |
| | IOD | 0.181336 | 0.03559 | 0.110289 | 1 | | |
| | NAO | 0.072 | 0.021 | -0.070 | 0.056 | 1 | -0.091 |
| | AMO | -0.081 | 0.040 | 0.031 | -0.0127 | -0.091 | 1 |

**Bold indicates the correlation values are significantly different from zero at 95% confidence levels.**

**Table 3 Summarized results from wavelet analyses**

| Station Name | Wavelet Periodicities | | | | | |
|---|---|---|---|---|---|---|
| | < 8 months | 8-16 months | 16-32 months | 32–64 Months | 64–128 Months | >128 months |
| Ajmer | - | in | - | - | in | in |
| Aligarh | hp | in | -- | in | yes | yes |
| Amarvati | yes | yes | -- | yes | hp-n | hp |
| Bandhra | - | yes | in | - | in | - |
| Bangalore | - | - | hp-n | hp-n | - | In |
| Bharauch | | in | - | - | hp | hp |
| Bikaner | - | In | - | - | in | in |
| Chennai | | yes | -- | in | hp | hp |
| Delhi | hp | in | -- | in | yes | yes |
| Dibrugarh | - | in | - | - | in | in |
| Gandhinagar | hp | in | in | in | yes | yes |
| Garhwal | in | In | - | - | In | hp |
| Guna | - | In | - | - | in | hp |
| Hydearbad | - | yes | - | - | in | - |
| Idukki | - | in | hp | - | - | - |
| Jabalpur | yes | in | -- | yes | yes | hp |
| Jaipur | - | in | - | - | hp | hp |
| Jalgoan | - | - | - | in | hp | Hp-n |
| Jammu | hp-n | yes | in | in | hp | hp |
| Jharsguda | - | in | - | - | hp | hp |
| Kodugu | - | Yes | - | - | in | in |
| Kolkata | -- | yes | -- | yes | yes | yes |
| Madurai | - | in | in | | in | - |
| North 24 parganas | - | in | - | - | hp | hp |
| Puri | -- | Yes | -- | yes | hp-n | yes |
| Ramnad | - | Yes | in | - | - | - |
| Shillong | - | in | - | - | in | in |
| Varanasi | - | - | - | in | in | in |
| Vizag | -- | Yes | -- | in | hp | hp |
| Wardha | yes | In | -- | yes | hp-n | hp |

yes: high power and significant, in: intermittent and significant, hp: high power detected but not significant at p < 0.05; hp-n: intermittent high power and not significant ; -- not significant

**Table 4 Correlation between the different rainfall and the climate indices at different scales.**

| Station | IOD | | | | | | | Nino 3.4 | | | | | | |
|---|---|---|---|---|---|---|---|---|---|---|---|---|---|---|
| | 0-2 | 2-4 | 4-8 | 8-16 months | 16-32 | 32-64 | 64-128 | 0-2 | 2-4 | 4-8 | 8-16 months | 16-32 | 32-64 | 64-128 |
| Ajmer | -0.07 | -0.09 | 0.05 | -0.12 | -0.28 | -0.06 | **0.20** | -0.02 | -0.02 | -0.03 | **-0.28** | -0.27 | **-0.45** | -0.15 |
| Aligarh | -0.01 | -0.09 | 0.03 | 0.19 | **-0.35** | -0.12 | 0.07 | 0.01 | 0.03 | 0.07 | **-0.22** | **-0.29** | **-0.46** | **-0.35** |
| Amarvati | -0.05 | -0.08 | -0.01 | 0.07 | -0.16 | **-0.29** | -0.03 | 0.06 | 0.09 | 0.11 | **-0.34** | -0.29 | **-0.59** | **-0.33** |
| Bandhra | 0.03 | -0.09 | -0.01 | 0.00 | -0.12 | -0.10 | **0.46** | 0.05 | 0.07 | 0.11 | **-0.24** | -0.05 | -0.31 | 0.17 |
| Bangalore | 0.01 | 0.02 | -0.02 | -0.06 | 0.01 | -0.03 | **-0.32** | 0.09 | 0.04 | 0.07 | **-0.33** | -0.15 | **-0.29** | -0.11 |
| Bharauch | -0.07 | 0.03 | 0.05 | -0.09 | -0.12 | 0.04 | **0.37** | 0.01 | 0.08 | 0.10 | **-0.31** | -0.31 | 0.11 | 0.05 |
| Bikaner | -0.05 | -0.08 | 0.00 | 0.00 | -0.11 | -0.07 | **0.10** | 0.02 | 0.09 | 0.09 | **-0.32** | -0.33 | **-0.49** | 0.02 |
| Chennai | -0.09 | 0.01 | **-0.27** | 0.04 | 0.10 | 0.14 | 0.09 | 0.04 | 0.02 | 0.02 | **-0.15** | 0.05 | 0.13 | **0.32** |
| Delhi | -0.02 | -0.09 | 0.04 | **0.21** | -0.25 | -0.03 | -0.03 | 0.01 | 0.00 | 0.07 | -0.23 | **-0.13** | **-0.37** | **-0.43** |
| Dibrugarh | 0.05 | 0.04 | 0.14 | 0.18 | 0.13 | 0.06 | **-0.51** | -0.02 | 0.11 | 0.01 | **-0.18** | -0.09 | -0.13 | -0.05 |
| Gandhinagar | 0.10 | 0.07 | 0.02 | 0.27 | -0.14 | -0.12 | **0.37** | 0.01 | 0.05 | 0.09 | **-0.30** | -0.26 | -0.10 | 0.20 |
| Garhwal | -0.01 | -0.10 | -0.01 | 0.01 | **-0.22** | -0.07 | 0.04 | 0.01 | 0.02 | 0.08 | **-0.23** | -0.07 | **-0.50** | **-0.30** |
| Guna | 0.02 | -0.05 | -0.01 | -0.01 | -0.05 | 0.15 | **-0.25** | 0.01 | 0.02 | 0.08 | 0.07 | **-0.25** | -0.03 | -0.07 |
| Hydearbad | -0.03 | 0.03 | -0.20 | 0.08 | 0.12 | 0.01 | **0.13** | 0.08 | -0.01 | 0.02 | 0.05 | -0.03 | 0.01 | **0.24** |
| Idukki | -0.06 | 0.12 | 0.03 | 0.04 | -0.16 | -0.09 | **0.18** | -0.01 | 0.03 | 0.05 | **-0.22** | -0.24 | **-0.47** | **-0.18** |
| Jabalpur | 0.01 | -0.09 | 0.00 | 0.00 | **-0.22** | **-0.60** | **0.28** | 0.06 | 0.05 | 0.08 | **-0.19** | -0.09 | **-0.57** | 0.02 |
| Jaipur | -0.03 | -0.09 | 0.05 | -0.13 | -0.31 | -0.13 | **0.22** | 0.04 | 0.01 | 0.05 | **-0.31** | -0.33 | **-0.62** | -0.07 |
| Jalgoan | 0.02 | 0.14 | 0.02 | 0.04 | **-0.29** | **-0.33** | -0.13 | 0.00 | 0.04 | 0.08 | **-0.31** | **-0.39** | **-0.62** | **0.19** |
| Jammu | -0.11 | -0.12 | 0.06 | 0.01 | 0.03 | -0.04 | **0.18** | 0.00 | 0.01 | 0.01 | **-0.20** | -0.13 | **-0.38** | -0.19 |
| Jharsguda | 0.08 | -0.10 | 0.02 | 0.04 | **-0.29** | -0.22 | **0.11** | 0.02 | 0.03 | 0.08 | 0.00 | -0.03 | -0.08 | -0.03 |
| Kodugu | -0.10 | 0.01 | 0.07 | 0.08 | -0.13 | 0.08 | **0.27** | 0.02 | 0.18 | 0.06 | **-0.20** | -0.19 | -0.18 | -0.11 |
| Kolkata | -0.03 | -0.06 | -0.01 | 0.00 | -0.09 | 0.10 | -0.11 | -0.02 | 0.01 | 0.08 | -0.05 | -0.25 | -0.04 | 0.00 |
| Madurai | 0.02 | 0.02 | 0.01 | -0.12 | 0.02 | -0.16 | **-0.19** | 0.07 | 0.00 | 0.02 | **-0.12** | -0.09 | -0.18 | -0.14 |
| North 24 parganas | -0.05 | 0.11 | 0.04 | 0.09 | -0.20 | -0.13 | **0.21** | -0.04 | 0.00 | 0.06 | **-0.33** | -0.29 | **-0.62** | -0.05 |
| Puri | -0.09 | 0.14 | 0.04 | 0.00 | **-0.28** | -0.30 | 0.09 | 0.00 | 0.04 | 0.09 | **-0.15** | -0.20 | **-0.34** | **0.36** |
| Ramnad | -0.05 | 0.04 | 0.05 | -0.16 | -0.07 | **-0.20** | **-0.17** | 0.07 | 0.12 | 0.01 | **-0.12** | -0.06 | -0.21 | -0.12 |
| Shillong | 0.09 | 0.16 | 0.16 | 0.16 | 0.06 | -0.03 | **-0.57** | 0.07 | 0.18 | 0.01 | 0.02 | -0.27 | -0.06 | -0.03 |
| Varanasi | -0.01 | -0.13 | -0.03 | 0.19 | **-0.22** | **-0.36** | **0.40** | 0.04 | 0.01 | 0.07 | 0.00 | -0.20 | **-0.34** | **-0.14** |
| Vizag | -0.07 | 0.15 | 0.08 | 0.15 | -0.25 | -0.08 | 0.14 | 0.00 | 0.05 | 0.11 | **-0.23** | 0.00 | -0.14 | **0.51** |
| Wardha | -0.02 | -0.11 | 0.01 | 0.01 | -0.08 | -0.11 | **0.19** | 0.04 | 0.06 | 0.11 | **-0.35** | -0.19 | **-0.55** | -0.11 |